\documentclass[conference]{IEEEtran}
\IEEEoverridecommandlockouts
\usepackage{cite}
\usepackage{amsmath,amssymb,amsfonts}
\usepackage{graphicx}
\usepackage{subcaption}
\usepackage{textcomp}
\usepackage{xcolor}
\usepackage{url}
\usepackage{textcomp}
\usepackage{xcolor}
\usepackage{booktabs}
\usepackage{amsmath}
\usepackage{adjustbox}
\usepackage{multirow}
\usepackage{svg}
\usepackage{caption}
\usepackage{ifthen}
\usepackage{amsfonts}
\usepackage{algorithm}
\usepackage{algorithmicx}
\usepackage{algpseudocode}
\usepackage{newtx}

\def\BibTeX{{\rm B\kern-.05em{\sc i\kern-.025em b}\kern-.08em
    T\kern-.1667em\lower.7ex\hbox{E}\kern-.125emX}}
\begin{document}

\title{NegBLEURT Forest: Leveraging Inconsistencies for Detecting Jailbreak Attacks}

\author{\IEEEauthorblockN{1\textsuperscript{st} Lama Sleem}
\IEEEauthorblockA{\textit{University of Luxembourg} \\
%\textit{name of organization (of Aff.)}\\
Luxembourg, Luxembourg \\
lama.sleem@uni.lu}
\\
\IEEEauthorblockN{4\textsuperscript{th} Nathan Foucher}
\IEEEauthorblockA{\textit{Institut National Polytechnique de Toulouse} \\
%\textit{name of organization (of Aff.)}\\
Toulouse, France \\
nathan.foucher@etu.toulouse-inp.fr}
\and

\IEEEauthorblockN{2\textsuperscript{nd} Jerome Francois}
\IEEEauthorblockA{\textit{University of Luxembourg} \\
%\textit{name of organization (of Aff.)}\\
Luxembourg, Luxembourg\\
jerome.francois@uni.lu}
\\
\IEEEauthorblockN{5\textsuperscript{th} Niccolo Gentile}
\IEEEauthorblockA{\textit{Foyer S.A.} \\
%\textit{name of organization (of Aff.)}\\
Leudelange, Luxembourg \\
niccolo.gentille@foyer.lu}

\and
\IEEEauthorblockN{3\textsuperscript{rd} Lujun Li}
\IEEEauthorblockA{\textit{University of Luxembourg} \\
%\textit{name of organization (of Aff.)}\\
Luxembourg, Luxembourg \\
lujun.li@uni.lu}
\\
\IEEEauthorblockN{6\textsuperscript{th} Radu State}
\IEEEauthorblockA{\textit{University of Luxembourg} \\
%\textit{name of organization (of Aff.)}\\
Luxembourg, Luxembourg \\
radu.state@uni.lu}
}

\maketitle

\begin{abstract}
Jailbreak attacks designed to bypass safety mechanisms pose a serious threat by prompting LLMs to generate harmful or inappropriate content, despite alignment with ethical guidelines. Crafting universal filtering rules remains difficult due to their inherent dependence on specific contexts. To address these challenges without relying on threshold calibration or model fine-tuning, this work introduces a semantic consistency analysis between successful and unsuccessful responses, demonstrating that a negation-aware scoring approach captures meaningful patterns. Building on this insight, a novel detection framework called NegBLEURT Forest is proposed to evaluate the degree of alignment between outputs elicited by adversarial prompts and expected safe behaviors. It identifies anomalous responses using the Isolation Forest algorithm, enabling reliable jailbreak detection. Experimental results show that the proposed method consistently achieves top-tier performance, ranking first or second in accuracy across diverse models using the crafted dataset, while competing approaches exhibit notable sensitivity to model and data variations\footnote{\scriptsize More results are available.  https://github.com/DobricLilujun/jailbreaktester}.
\end{abstract}


\section{Introduction}
\label{sec:intro}

Large Language Models (LLMs) are powerful neural networks with large parameter sizes and strong in-context learning capabilities, widely used for tasks such as summarization, text completion, and answering questions~\cite{hadi2023survey,openai2023gpt4v,kasneci2023chatgpt,zhao2023survey}. Popular models include GPT-3~\cite{mann2020language}, GPT-4~\cite{achiam2023gpt}, and Llama~\cite{touvron2023llama}, which are typically accessed through APIs or web interfaces. These models are pretrained on large-scale corpora encompassing extensive world knowledge, which may also include harmful or illegal content. However, this broad accessibility and extensive knowledge base also expose these models to various cyber threats, such as prompt-based attacks, which can manipulate model behaviors and potentially compromise system security~\cite{perez2022ignore}. One of the most common threats is the jailbreak attack, which aims to bypass safety mechanisms and induce the model to generate harmful or illegal content according to local laws and regulations~\cite{chao2023jailbreaking,zou2023universal}. 

State-of-the-art methods, such as SmoothLLM~\cite{robey2023smoothllm}, detect jailbreak attacks by introducing controlled perturbations to input prompts and aggregating the corresponding model outputs through majority voting. This process estimates the likelihood of different responses and determines whether the attack succeeds on a given model. Reproducing prior methods such as JailGuard~\cite{zhang2023mutation} is not always straightforward, as implementation details and unclear threshold choices can lead to variations in results, even on identical datasets. These issues reflect broader reproducibility and comparability problems in the field. In conclusion, this work therefore explores two key questions: \textbf{RQ1: What are the principal semantic differences between successful and failed attacks, and is it feasible to detect them without relying on a predefined threshold?} \textbf{RQ2: How can we distinguish semantically successful attacks from unsuccessful ones using one robust and generalizable framework?} To address these questions, we conducted extensive experiments to investigate semantic inconsistency using various embedding-based and negation-oriented metrics. Based on our findings, we propose a novel framework, NegBLEURT Forest, for effective and reliable detection of successful jailbreak attacks. This work is organized as follows: Section~\ref{sec:related_work} reviews existing jailbreak attacks and defense mechanisms. Section~\ref{sec:approach} presents an analysis of semantic inconsistency between successful and unsuccessful attacks using embedding-based and negation-aware metrics. Section~\ref{sec:Method Derived} introduces the proposed NegBLEURT Forest framework, along with its experimental evaluation and analysis. Finally, Section~\ref{sec:conc} concludes the paper and discusses the limitations of the proposed approach.

\section{Related Work}
\label{sec:related_work}

Jailbreaking refers to the act of bypassing a model’s safeguards so that a harmful prompt (vanilla prompt) \( P \), which would normally trigger a benign response \( R_{\text{benign}} \), instead elicits a harmful response \( R_{\text{harm}} \) from the target model \( M \) after modifying the original \( P \) to an adversarial prompt \( P' \)~\cite{rebedea-etal-2025-guardrails}. These attacks fall into two main categories: conflicting goals, which force models to choose between safe and harmful responses (e.g., GCG and AutoDAN~\cite{zou2023universal}), and generalization mismatches, which exploit the gap between pretraining and safety fine-tuning~\cite{wang2018efficient}. Defenses are typically categorized as preprocessing (e.g. smoothing or detection~\cite{ji2024defending, robey2023smoothllm}) and postprocessing (e.g. output filtering~\cite{pisano2023bergeron, phute2023llm}). Both methods demonstrate strong effectiveness but also exhibit notable limitations. Pre-processing approaches rely on predefined threshold values to guide the final classification decisions when distinguishing between safe and harmful prompts; however, these thresholds are often selected without rigorous justification and typically address only one or two specific attack types, which restricts their generalization. Post-processing approaches require adapting the model through filter tuning, a method that is time consuming and resource intensive. In addition, the reliability of these methods is not always guaranteed, particularly when external LLMs are used for responses. Such models are typically closed-source and accessible only through APIs, making them impossible to further train or fine-tune for specific defense. Numerous studies have introduced jailbreak attack detection approaches within these categories. For example, SmoothLLM~\cite{robey2023smoothllm}, JailGuard~\cite{zhang2023mutation}, LlamaGuard~\cite{inan2023llama}, Perplexity-based detection methods~\cite{alon2023detecting}, and various defense strategies~\cite{metzen2017detecting, liu2022complex} have been developed to detect attacks and strengthen AI security. These techniques primarily operate at the response level and the algorithm's final decisions typically rely on predefined thresholds or specific word labels, such as "I cannot," which are neither generalizable nor stable.

\section{Consistency Analysis}
\label{sec:approach}

Inconsistency between outputs is commonly assessed via embedding-based approaches, wherein semantic information is mapped into a vector space using methods such as embedding transformers. However, these embeddings may sometimes fail to accurately capture critical information related to affirmation or negation \cite{li2025exploringimpacttemperaturelarge}. In this section, we systematically examine the key differences between successful and unsuccessful attacks, focusing on both embedding similarity and scoring that is aware of negations. Specifically, we employ NegBLEURT~\cite{anschutz2023not} to provide a negation sensitive score and compare its effectiveness with analyses based on cosine similarity. This methodological comparison enables a comprehensive evaluation of the consistency, stability, and sensitivity of model outputs under various adversarial attacks, including cases involving subtle prompt perturbations. 

\subsection{Datasets}

Building on the JailbreakBench datasets\cite{chao2024jailbreakbench}, introduced in \cite{andriushchenko2024jailbreaking}, as well as JailbreakV-28K~\cite{luo2024jailbreakv}, we created a carefully curated and manually labeled dataset containing 161 original harmful prompts. Although JailbreakBench serves as an informative starting point, manual labeling is necessary to accurately determine the outcome of each prompt. Manual validation reveals that many prompts labeled as successful in JailbreakV-28K do not consistently produce jailbreaks, highlighting the need for human review. The final dataset represents a balanced combination of both sources, with all entries manually validated for accuracy. 

Different perturbations are then applied  to examine the impact of changes in responses drawing inspiration from SmoothLLM. The semantic similarity of the resulting responses is analyzed to gain deeper insight into how such perturbations affect the model's behavior at the semantic level. Three techniques are used: Insert Perturbation, which randomly adds contextually fitting words or phrases to the prompt to check if the meaning stays consistent; Patch Perturbation, which replaces certain words or phrases with alternatives while keeping the sentence structure to see how the model adapts; and Swap Perturbation, which changes the order of words or phrases to test the model’s ability to understand the prompt despite word rearrangement. To better detect inconsistencies, we apply six perturbation levels ($1\%$, $3\%$, $5\%$, $10\%$, $15\%$ and $25\%$), generating 10 variations at each level. For each perturbed prompt, we generated 10 different responses from the model to get a reliable sample for evaluating how consistently the model reacts to small input changes. This also helped us create a ground truth for further analysis. Having multiple responses per prompt allows us to measure response consistency, which shows how similar the responses are to each other and reflects the model’s defense stability. 

\subsection{Experiments}

To analyze inconsistencies between generated responses, we examine two types of similarity measures: \textbf{cosine similarity} and the \textbf{NegBLEURT score}. Specifically, the objective is to determine whether notable changes occur in the generated responses when the original prompt is slightly modified, and to assess whether these two similarity metrics can effectively distinguish between successful and unsuccessful responses. The underlying hypothesis is that prompts engineered as attacks will yield responses exhibiting significant variation when the attack is successful. In contrast, prompts that consistently fail or succeed will produce highly similar responses, predominantly reflecting either harmful or safe content. Therefore, when a high degree of inconsistency is observed across responses, the corresponding prompt is likely to represent a successful jailbreak attack.

To elaborate more precisely, given an initial prompt $P_0$, we generate a set of $n$ responses $\mathcal{R} = \{ R_1, R_2, \ldots, R_n \}$, where $n = 10$ in our experiments. For any pair of distinct responses $R_i$ and $R_j$ ($1 \leq i, j \leq n$, $i \neq j$), their similarity is denoted as $S(R_i, R_j)$, measured using a designated metric such as cosine similarity or NegBLEURT. For each response $R_i$, we define its average similarity to all other responses (i.e., its "1-vs-all" consistency) as:

\begin{equation}
\mu_{\max} = \max_{1 \leq i \leq n} \left( \frac{1}{n - 1} \sum_{\substack{j=1 \\ j \neq i}}^{n} S(R_i, R_j) \right)
\end{equation}

Here, $\mu_{\max}$ is defined as the maximum of the average scores across all samples. We denote $\mu_{\max}$ computed with NegBLEURT as $\mu_{\max}(\text{Neg})$, and $\mu_{\max}$ computed with semantic embeddings using cosine similarity as $\mu_{\max}(\text{Cos})$.

\begin{figure}[!htbp]
    \centering
    \begin{subfigure}[b]{0.90\linewidth}
        \centering
        \includegraphics[width=\linewidth]{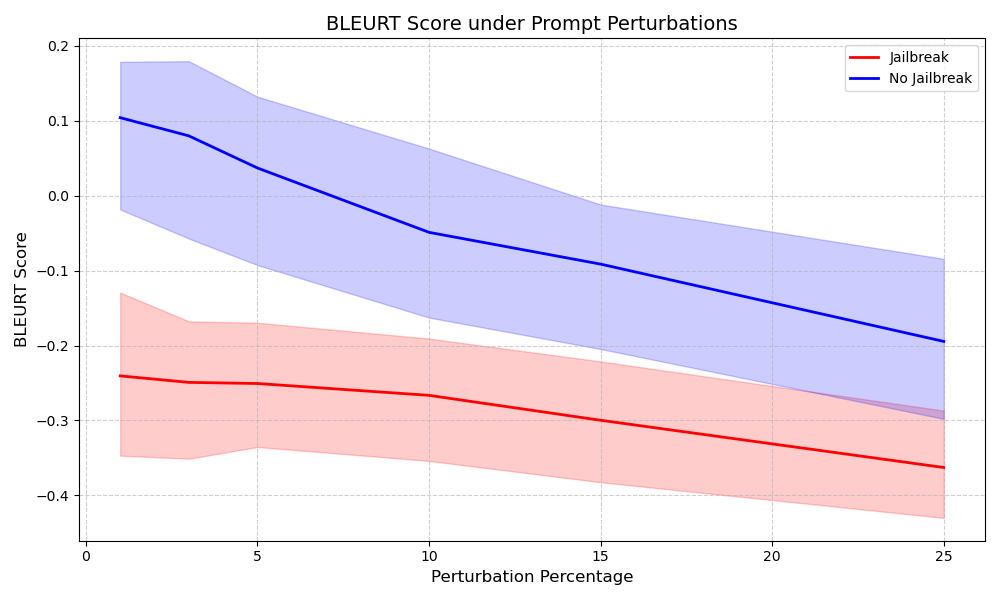}
        \caption{$\mu_{\max}(\text{Neg})$}
        \label{fig:Neg}
    \end{subfigure}
    
    \vspace{0.5em}
    
    \begin{subfigure}[b]{0.90\linewidth}
        \centering
        \includegraphics[width=\linewidth]{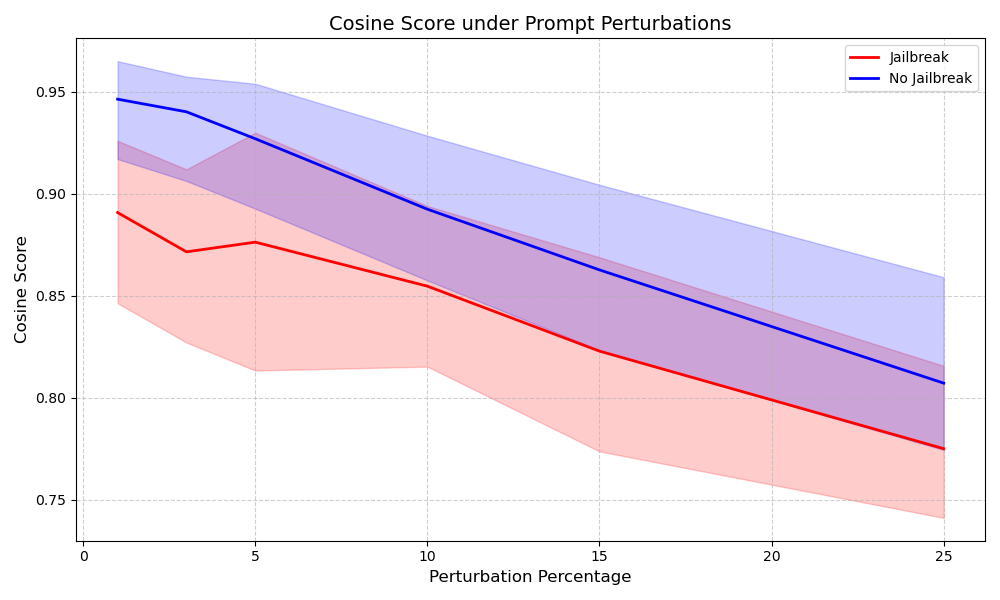}
        \caption{$\mu_{\max}(\text{Cos})$}
        \label{fig:embed}
    \end{subfigure}

    \caption{Inconsistency variation across datasets with different perturbation levels. Each curve shows the test set average of $\mu_{\max}$, with shaded areas indicating the interquartile range (25\%--75\%). Blue curves indicate consistent outputs (unsuccessful attacks), while red curves indicate changed outputs (successful attacks).}
    \label{fig:llama_negbleurt_cosine}
\end{figure}

Fig. \ref{fig:llama_negbleurt_cosine} shows that from a statistical perspective, semantic cosine similarity cannot effectively distinguish between successful and unsuccessful responses, as evidenced by the substantial overlap between the red and blue regions corresponding to $\mu_{\max}(\text{Cos})$. In contrast, when we employ the NegBLEURT score—a negation-oriented metric—it is able to statistically discern the difference in similarity between good and poor responses much more effectively, as illustrated by the trend in Figure \ref{fig:Neg}. It is also observed that as the perturbation ratio increases, especially when it exceeds 25\%, the uncertainty in the model's responses grows significantly. Semantic inconsistency then manifests not only in refusals or negations, but also in the manner of answering questions and in the model's interpretation of the prompts. 

Consequently, relying solely on NegBLEURT to distinguish between successful and failed attacks is insufficient, since there remains a nontrivial overlap between the two regions, as illustrated in Figure~\ref{fig:Neg} when the perturbation rate reaches 20\%, within the 25--75\% quantile range. Based on these observations, we derive several key insights. NegBLEURT proves to be an effective means for distinguishing semantic consistency among responses, and such inconsistencies are salient features of jailbreak attacks. An excessively low value of $\mu_{\max}(\text{Neg})$ serves as a strong indicator of potential attacks. However, excessive perturbation reduces the ability to distinguish between attack and non-attack responses, rendering the sole use of the NegBLEURT similarity score as a criterion unreliable. To address these challenges and to leverage main findings in this section, we propose NegBLEURT Forest—a method for detecting jailbreak attacks by exploiting semantic consistency as measured by NegBLEURT score and other embeddings.

\section{Method Derived}
\label{sec:Method Derived}

\subsection{Proposed Framework}

\begin{figure*}[h]
    \centering
    \includegraphics[width=0.9\textwidth]{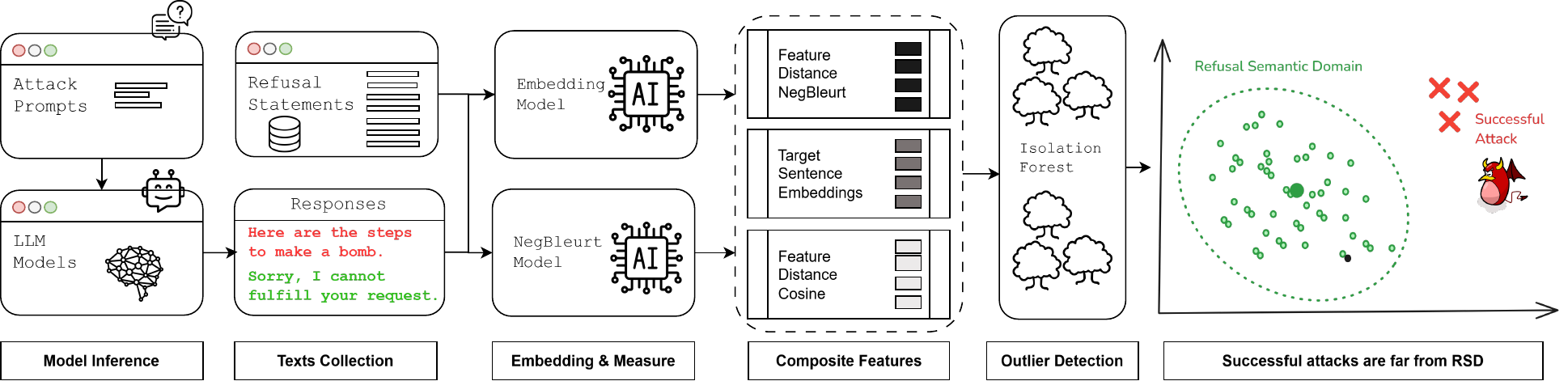}
    \caption{The proposed framework for detecting successful harmful prompt attacks using $\mathbb{RSD}$ and Isolation Forest.}
    \label{fig:svg_example}
\end{figure*}

We propose the NegBLEURT Forest jailbreak detection framework, as illustrated in Figure~\ref{fig:svg_example}, which is designed to systematically assess the robustness of LLMs in rejecting prompts that include harmful requests, illicit content, or violate predefined safety protocols. Building on the previously introduced NegBLEURT as a metric to detect semantic inconsistency, the model accounts for discrepancies between LLM responses and the refusal-related semantic domain. By also integrating embedding-based representations, a composite feature space is formed that improves the scalability and generalizability of the proposed detection framework.

\subsubsection{Refusal Semantic Domain ($\mathbb{RSD}$)}

\begin{table}[!htbp]
\centering
\caption{Typical model responses to harmful queries, categorized by refusal type.}
\label{tab:refusal-types}
\begin{adjustbox}{width=0.98\linewidth}
\begin{tabular}{@{}ll@{}}
\toprule
\textbf{Refusal Type} & \textbf{Example} \\
\midrule
\textbf{Direct Refusal} & I cannot fulfill your request. \\
\textbf{Refusal + Apology} & 
\begin{tabular}[t]{@{}l@{}}I apologize, but I'm a large
language model, I cannot provide \\
you with that information.\end{tabular} \\
\textbf{Refusal + Disclaimer} & 
\begin{tabular}[t]{@{}l@{}}I must inform you that the
prompt you provided contains \\ harmful content, and I cannot assist
with that.\end{tabular} \\
\bottomrule
\end{tabular}
\end{adjustbox}
\end{table}

Based on a large collection of both successful and failed jailbreak attempts and their corresponding responses, we observe that when models refuse harmful user queries, their outputs generally fall into three categories: (1) direct refusal, (2) refusal accompanied by an apology, and (3) refusal with a disclaimer. These response patterns are exemplified in Table~\ref{tab:refusal-types}, where refusal to answer constitutes the key semantic feature of these outputs. Let $\mathcal{D}_{\text{rej}}$ denote a subset of rejection-related utterances (rejection corpus) collected for analysis. 

We define $\mathbb{RSD}$ as a subset of $\mathcal{S}$, where $\mathcal{S}$ is the space of all possible sentences in natural language such that $\mathbb{RSD}$ consists of sentences in $\mathcal{D}_{\text{rej}}$ whose length is between 15 and 20 tokens (inclusive). The range of sentence lengths between 15 and 20 tokens reflects the optimal compression rate and the information transmission efficiency of natural language\cite{borbely-kornai-2019-sentence}. Moreover, the distribution of English sentence lengths follows a modified Zipf-Mandelbrot distribution, whose parameter optimization points to a peak interval of 15–20 words\cite{Wordlength2004}. Empirical studies across various fields have confirmed the effectiveness of sentence lengths within this range, such as Martin Cutts, author of The Oxford Guide to Plain English~\cite{cutts2020oxford}, advises keeping the average sentence length between 15 and 20 words throughout a document. 

Here, $\text{len}(s)$ denotes the number of words in the sentence $s$. While $\mathbb{RSD}$ includes infinitely many texts, their semantics often overlap. Therefore, a finite set of related examples from $\mathcal{D}_{\text{rej}}$ can effectively represent the $\mathbb{RSD}$ in semantic space. We assume a reliable model should consistently reject unsafe or harmful prompts. Successful jailbreaks may appear as direct answers (e.g., to “How to make a bomb”) or as responses that initially refuse but later reveal harmful content. A response is considered robust if it aligns semantically and distributionally with the $\mathbb{RSD}$, showing no deviation from expected refusal behavior. Around 50 common refusal phrases were used, though this set is extendable for flexibility and improved robustness.

\subsubsection{Proposed Method}
In this framework, we treat $\mathcal{D}_{\text{rej}}$ as the set of reference responses, and the generated responses as the targets. These responses are transformed into high-dimensional vectors (embeddings) one sentence transformer. We then apply the K-Means algorithm to identify the semantic center of $\mathcal{D}_{\text{rej}}$. Specifically, the KMeans algorithm is only applied to reference embeddings $\mathbf{e}_{\text{ref}}$. In this setting, the number of clusters $k$ is set to 1, which means that the algorithm produces a single cluster center $\mathbf{c}_{\text{ref}}$, which serves as a representative summary of the reference responses:
\begin{equation}
\mathbf{c}_{\text{ref}} = \text{KMeans}(\mathbf{e}_{\text{ref}}, k=1)
\end{equation}

Next, we define the cosine similarity distance between embeddings. Specifically, we employ cosine similarity as the distance metric, given a text embedding $e_{tgt}$, 

\begin{equation}
D_{\text{emb}} (e_{tgt}, \mathbf{c}_{\text{ref}}) = \frac{e_{tgt} \cdot \mathbf{c}_{\text{ref}}}{\|e_{tgt}\| \, \|\mathbf{c}_{\text{ref}}\|}
\end{equation}

In addition, we define another type of distance, referred to as the \textit{NegBLEURT Score Distance}. NegBLEURT is a model designed to compare the semantic similarity between two texts. A higher score indicates stronger similarity---scores can potentially exceed 1 and are generally below 2. Conversely, dissimilar or contradictory pairs tend to yield scores below 0. Therefore, for each reference response in \( \mathcal{D}_{\text{rej}} \), we compute a NegBLEURT score with respect to the target response. 

\begin{equation}
D_{\text{Neg}} (e_{\text{tgt}}, \mathcal{E}_{\text{rej}}) = \left[ M_{\text{Neg}}(e_{\text{tgt}}, e_i) \right]_{i=1}^{N}
\end{equation}

\noindent
where \( e_{\text{tgt}} \) denotes the target embedding of response generated by the LLM, which may correspond to either a successful or an unsuccessful attempt, and \( \mathcal{E}_{\text{rej}} = \{e_1, e_2, \dots, e_N \} \). \( M_{\text{Neg}} \) denotes the NegBLEURT model as a function. We assume that the target response embedding \( e_{\text{tgt}} \in \mathbb{R}^{E \times 1} \), and the NegBLEURT Score Distance vector \( D_{\text{Neg}}(e_{\text{tgt}}, \mathcal{E}_{\text{rej}}) \in \mathbb{R}^{N \times 1} \). In contrast, the embedding-based distance between the target and reference response, \( D_{\text{emb}}(e_{\text{tgt}}, c_{\text{ref}}) \), is a scalar, i.e., \( \in \mathbb{R}^{1 \times 1} \). 

To ensure that these components contribute equally in the representation of the joint characteristics, we extend both \( D_{\text{Neg}} \) and \( D_{\text{emb}} \) to \( E \times 1 \), i.e. \( 768 \times 1 \) vectors through replication, denoted \( D_{\text{Neg}}' \) and \( D_{\text{emb}}' \), respectively. The complete feature representation for each item is then defined as:

\begin{equation}
F(e_{\text{tgt}}, \mathbb{RSD}) = \left[ e_{\text{tgt}} \;\middle|\; D_{\text{Neg}}' \;\middle|\; D_{\text{emb}}' \right]
\label{eq:composite_feature_vector}
\end{equation}
where $F(e_{\text{tgt}}, \mathbb{RSD}) \in \mathbb{R}^{3E \times 1}$. In this case, the features not only encode semantic information, but also incorporate the similarity distance between the semantics and the $\mathbb{RSD}$, as well as the distance to Negation.
\subsubsection{Outlier Detection (Iso-Forest)}
Isolation Forest \cite{iso_forest} is an anomaly detection method that works by isolating samples in a dataset using random partitioning. Intuitively, anomalies are easier to isolate and thus have shorter path lengths in the isolation trees. The anomaly score for each sample is computed based on the average path length required to isolate it, normalized by the expected path length for a given sample size. Samples with higher anomaly scores are considered more likely to be outliers. To detect anomalies, we select samples with scores above a certain threshold, determined by a contamination rate \(\alpha\). In this work, we set \(\alpha = \frac{1}{N+1}\), meaning we expect to find exactly one anomaly among \(N\) samples.

\subsubsection{Extraction Framework}

The model output exhibits a certain degree of randomness and, depending on the input, may occasionally produce apologies or refusals. In particular, the responses vary according to the specific requirements of the input. Although the general semantics may resemble $\mathbb{RSD}$, the embeddings extracted by the model capture a broader spectrum of semantic information. Consequently, while the output may contain elements of refusal, it also encompasses other semantic meanings, which can lead the isolation forest algorithm to identify the output as an outlier. To address this issue, we employ an extraction algorithm to obtain the core attitudinal information, as described in Algorithm~\ref{alg:salient_sentence_extraction}. In this study, the zero-shot classifier is implemented in an unsupervised manner using the pre-trained model \texttt{facebook/bart-large-mnli}. Specifically, only a set of candidate labels \(\mathcal{L}\) is defined, and the model subsequently computes a classification score for each label based on the given input. The main objective of this algorithm is to extract the emotionally expressive parts of a sentence and truncate it to an appropriate length for subsequent analysis and recognition.

\begin{algorithm}
\caption{Extraction of Salient Sentence}
\label{alg:salient_sentence_extraction}
\begin{algorithmic}[1]
\Require Text $T$; Zero-shot classifier $C$; Label set $\mathcal{L} = \{\text{refusal}, \text{apology}, \text{informative}\}$
\Ensure Salient sentence $S^*$

\State Split $T$ into $N$ sentences: $\{s_1, s_2, \dots, s_N\}$
\For{$i = 1$ to $N$}
    \State Compute scores: $\mathbf{p}_i \gets C(s_i, \mathcal{L})$
    \State $\ell_i \gets \arg\max_{\ell \in \mathcal{L}} \mathbf{p}_i[\ell]$
\EndFor

\State Define $\mathcal{L}_{\text{emo}} \gets \{\text{refusal}, \text{apology}\}$
\State $\mathcal{S}_{\text{emo}} \gets \{s_i \mid \ell_i \in \mathcal{L}_{\text{emo}}\}$

\If{$\mathcal{S}_{\text{emo}} \neq \emptyset$}
    \State $S^* \gets$ sentence in $\mathcal{S}_{\text{emo}}$ with highest emotional score
\Else
    \State $S^* \gets s_1$ Fallback to the first sentence
\EndIf

\If{the length of $S^*$ is outside the range [15, 20]}
    \State Trim $S^*$ by semantic segmentation and keep the segment with highest emotional score
\EndIf

\State \Return $S^*$
\end{algorithmic}
\end{algorithm}

\subsubsection{Methodology Overview}

In summary, as illustrated in Figure~\ref{fig:svg_example}, given a harmful trigger, the model first generates a response through inference. This response is then subjected to an extraction algorithm to identify the most salient sentences. Subsequently, a composite feature vector is computed, which encapsulates two types of semantic distances—one comparing the \(\mathbb{RSD}\) and the other within the embedding space. Finally, Isolation Forest is employed to detect outliers based on this feature representation.

\begin{equation}
J = I\big(F(E(M(x)),\mathbb{RSD})\big)
\end{equation}

\noindent
where \(J\) denotes the jailbreak result, \(M\) the LLM model, \(E\) the extraction function, \(F\) the feature computation shown in Equation \ref{eq:composite_feature_vector}, \(I\) the Iso-Forest outlier detection and \(x\) the input harmful prompt.

\subsection{Experiments}

NegBLEURT Forest framework effectively addresses the issue of inconsistent output caused by the random nature of model responses. Instead of relying on explicitly defined refusal strings, it introduces an $\mathbb{RSD}$-based outlier detection mechanism, eliminating the need to manually specify classification thresholds. We adopt standard evaluation metrics including accuracy, precision, recall, and F1 score to assess performance. After perturbing the prompts,  evaluation on two different models using this expanded dataset is carried out. To validate that it outperforms the SOTA, we evaluated String-based Text Classification, Perplexity-guided Classification, Smoothed Language Model Classification, and the JailGuard method in the same test set, obtaining the results shown in Table \ref{tab:performance_results}. 

Furthermore, a comprehensive evaluation was made, encompassing not only its overall performance but also a series of ablation studies designed to systematically quantify the individual contributions of its constituent components to the model’s detection efficacy. Specifically, the investigation involved the exclusion of the Extraction Framework (denoted as Model w/o Extraction) and the isolated removal of critical elements within the NegBLEURT distance calculation (Model w/o NegBLEURT Distance) and the Embeddings (Model w/o Embeddings). Additionally, the study examined the effect of employing alternative embedding models—specifically, the \texttt{msmarco-distilbert-base-tas-b} model (Model with Another Model)—on detection performance. Finally, the robustness of the framework was assessed by evaluating a variant in which the representational dimensionality of the size of $\mathbb{RSD}$ was reduced by half (Model with Half Reference).

For the generation configuration of the LLMs, we adopted a consistent set of parameters across all experiments. The temperature was fixed at 1.0 to balance creativity and stability, the maximum output length was limited to 256 tokens, and the top-p sampling parameter was set to 0.9 to control diversity. No frequency penalty was applied, ensuring that repeated tokens were not artificially discouraged and the random seed was fixed at 47 to guarantee reproducibility.

\subsection{Results and Discussion}

\subsubsection{Detection Results}

\begin{table*}[!htbp]
\centering
\begin{adjustbox}{width=\textwidth}
\begin{tabular}{llcccccccccccccccc}
\toprule
\multirow{2}{*}{Methods} &
  \multirow{2}{*}{Models} &
  \multicolumn{4}{c}{Original Dataset (OD)} &
  \multicolumn{4}{c}{OD Patch Perturbation 25\%} &
  \multicolumn{4}{c}{OD Insert Perturbation 25\%} &
  \multicolumn{4}{c}{OD Swap Perturbation 25\%} \\
  \cmidrule(lr){3-18}
 &
   &
  \textbf{Accuracy} &
  \textbf{Precision} &
  \textbf{Recall} &
  \textbf{F1} &
  \textbf{Accuracy} &
  \textbf{Precision} &
  \textbf{Recall} &
  \textbf{F1} &
  \textbf{Accuracy} &
  \textbf{Precision} &
  \textbf{Recall} &
  \textbf{F1} &
  \textbf{Accuracy} &
  \textbf{Precision} &
  \textbf{Recall} &
  \textbf{F1} \\
  \midrule
\multirow{5}{*}{\textbf{Llama-2-7b-chat-hf}} &
  STR-CLS &
  0.435 &
  0.257 &
  0.122 &
  0.165 &
  0.863 &
  0.541 &
  0.800 &
  0.645 &
  0.857 &
  0.514 &
  0.750 &
  0.610 &
  0.913 &
  0.300 &
  \textbf{1.000} &
  0.462 \\
 &
  PPL-CLS &
  0.609 &
  0.867 &
  0.176 &
  0.292 &
  0.770 &
  0.167 &
  0.120 &
  0.140 &
  0.795 &
  0.200 &
  0.125 &
  0.144 &
  0.894 &
  0.077 &
  0.167 &
  0.105 \\
 &
  JailGuard &
  0.559 &
  0.667 &
  0.081 &
  0.145 &
  0.826 &
  0.333 &
  0.120 &
  0.177 &
  0.826 &
  0.333 &
  0.167 &
  0.222 &
  \textbf{0.919} &
  0.231 &
  0.500 &
  0.316 \\
 &
  SMLM-CLS &
  0.578 &
  \textbf{0.875} &
  0.095 &
  0.171 &
  0.839 &
  0.474 &
  0.360 &
  0.409 &
  0.820 &
  0.407 &
  0.458 &
  0.431 &
  \textbf{0.919} &
  0.111 &
  0.167 &
  0.133 \\
 &
  NegBleurtForest &
  \textbf{0.894} &
  0.817 &
  \textbf{1.000} &
  \textbf{0.899} &
  \textbf{0.870} &
  \textbf{0.692} &
  \textbf{0.878} &
  \textbf{0.774} &
  \textbf{0.870} &
  \textbf{0.673} &
  \textbf{0.897} &
  \textbf{0.769} &
  0.913 &
  \textbf{0.625} &
  0.750 &
  \textbf{0.682} \\
  \midrule
\multicolumn{1}{l}{\multirow{5}{*}{\textbf{Gemma-2-9b}}} &
  STR-CLS &
  0.851 &
  0.753 &
  0.939 &
  0.836 &
  0.776 &
  0.607 &
  0.944 &
  0.739 &
  0.857 &
  0.778 &
  0.927 &
  0.846 &
  0.683 &
  0.410 &
  0.944 &
  0.571 \\
\multicolumn{1}{l}{} &
  PPL-CLS &
  0.721 &
  0.667 &
  0.615 &
  0.640 &
  0.683 &
  0.517 &
  0.833 &
  0.638 &
  0.727 &
  0.700 &
  0.618 &
  0.656 &
  0.559 &
  0.289 &
  0.667 &
  0.403 \\
\multicolumn{1}{l}{} &
  JailGuard &
  0.752 &
  0.931 &
  0.415 &
  0.575 &
  0.696 &
  0.619 &
  0.241 &
  0.347 &
  0.671 &
  0.703 &
  0.382 &
  0.495 &
  0.677 &
  0.340 &
  0.472 &
  0.395 \\
\multicolumn{1}{l}{} &
  SMLM-CLS &
  \textbf{0.988} &
  \textbf{0.985} &
  \textbf{0.985} &
  \textbf{0.985} &
  0.795 &
  0.621 &
  \textbf{1.000} &
  0.766 &
  0.907 &
  0.844 &
  \textbf{0.956} &
  0.897 &
  0.603 &
  0.354 &
  0.944 &
  0.515 \\
\multicolumn{1}{l}{} &
  NegBleurtForest &
  0.901 &
  0.803 &
  1.000 &
  0.890 &
  \textbf{0.820} &
  \textbf{0.832} &
  0.859 &
  \textbf{0.845} &
  \textbf{0.907} &
  \textbf{0.878} &
  0.952 &
  \textbf{0.911} &
  \textbf{0.876} &
  \textbf{0.881} &
  \textbf{0.881} &
  \textbf{0.881} \\
\bottomrule
\end{tabular}
\end{adjustbox}
\caption{This table presents a comparative analysis of five classification approaches: STR-CLS (String-based Text Classification), PPL-CLS (Perplexity-guided Classification), SMLM-CLS (Smoothed Language Model Classification), JailGuard, and the proposed method, NegBLEURTForest. The evaluation is conducted on both the original clean dataset (OD) and a perturbed version containing 25\% noise derived from the OD. The results illustrate the robustness and effectiveness of each method under varying data conditions.
}
\label{tab:performance_results}
\end{table*}

As shown in Table \ref{tab:performance_results}, we observe that the model achieves the highest F1 scores in most cases, although SMLM-CLS performs relatively better on the Gemma model. It is worth highlighting that the method consistently achieves very high performance in all four test sets. However, despite SMLM-CLS achieving strong results in the OD dataset, its performance in the OD-SWAP is notably poor, significantly lower than NegBLEURT Forest’s 0.881. This further validates that the approach demonstrates greater generalizability, maintaining comparable high performance across different datasets, especially on responses generated by different models. It is also important to note that the performance of PPL-CLS is highly sensitive to the choice of the perplexity threshold. In this study, the threshold was selected to produce relatively high accuracy; however, its performance on the four data sets remains suboptimal, particularly in terms of the F1 score.

\subsubsection{Ablation Results}
It can be seen that this framework achieves high performance in both models in tests, particularly in terms of the F1 score, as shown in Table \ref{tab:full_results_ablation}. In addition, it was found that each component of the framework contributes positively to the overall performance of the model. For example, in the evaluation using Llama-2-7b-chat-hf, reducing the dimensionality of the $\mathbb{RSD}$ by half led to a notable performance degradation, with the F1 score dropping from 0.869 to 0.759. Furthermore, this demonstrates consistently strong performance across all tested models. When the Extraction Framework is removed, although relatively good results are maintained on the Gemma-2-9b model, the performance on Llama-2-7b-chat-hf deteriorates significantly, with an F1 score of only 0.726, significantly lower than the 0.869 achieved by the full model.

\begin{table}[ht]
\centering
\begin{adjustbox}{width=\columnwidth}
\begin{tabular}{clcccc}
\toprule
\multirow{2}{*}{\textbf{Methods}}            & \multicolumn{1}{c}{\multirow{2}{*}{\textbf{Model}}} & \multicolumn{4}{c}{\textbf{Full Dataset}}                      \\
                                             & \multicolumn{1}{c}{}                                & \textbf{ACC} & \textbf{Prec.} & \textbf{Rec.} & \textbf{F1}    \\
\midrule
\multirow{6}{*}{\textbf{Llama-2-7b-chat-hf}} & Base Framework                                      & 0.933        & 0.856          & 0.883         & \textbf{0.869} \\
                                            & Model w/o Extraction                                & 0.823        & 0.593          & 0.932         & 0.726          \\
                                            & Model w/o NegBleurt Distance                        & 0.888        & 0.821          & 0.710         & 0.762          \\
                                            & Model w/o Embeddings                                & 0.905        & 0.756          & 0.920         & 0.830          \\
                                            & Model with Half Reference                           & 0.849        & 0.635          & 0.944         & 0.759          \\
                                            & Model with Another Model                            & 0.904        & 0.798          & 0.827         & 0.812          \\
\midrule
\multirow{6}{*}{\textbf{Gemma-2-9b}}         & Base Framework                                      & 0.876        & 0.930          & 0.815         & \textbf{0.868} \\
                                            & Model w/o Extraction                                & 0.877        & 0.849          & 0.920         & 0.883          \\
                                            & Model w/o NegBleurt Distance                        & 0.800        & 0.926          & 0.653         & 0.767          \\
                                            & Model w/o Embeddings                                & 0.899        & 0.909          & 0.890         & 0.899          \\
                                            & Model with Half Reference                           & 0.873        & 0.842          & 0.920         & 0.879          \\
                                            & Model with Another Model                            & 0.800        & 0.945          & 0.639         & 0.762          \\
\bottomrule
\end{tabular}
\end{adjustbox}
\caption{Performance comparison of different models and configurations on the full dataset. We combined all data with 25\% perturbation from  main experiments with the metadata to construct a \(4 \times 161\) dataset, which we refer to as the Full Dataset.
}
\label{tab:full_results_ablation}
\end{table}
\section{Conclusion}
\label{sec:conc}
In this work, by systematically analyzing the responses of perturbed prompts,  using NegBLEURT and cosine similarity, it was found that NegBLEURT performs better in finding patterns of successful jailbreak which  emphasizes the role of negation to differentiate between successful and unsuccessful attacks. Building on insights from NegBLEURT and integrating Isolation Forest with a predefined $\mathbb{RSD}$ domain, the proposed method demonstrates strong performance and stability across the curated dataset. This framework is not only effective for the curated and manually labeled $\mathbb{RSD}$ related content but also demonstrates consistent performance across different models and different perturbed datasets. 

The core contribution of this work lies in addressing the increasing diversity of model refusal behaviors and enhancing the robustness of successful attack detection: as refusals become more varied, fixed criteria for determining whether a model rejects hazardous responses become less reliable. In contrast, the proposed NegBLEURT Forest provides an inherently semantic evaluation mechanism, assessing model behavior holistically without a predefined threshold rather than relying on simple similarity measures or pre-selected target words. Comparative analysis with existing approaches also confirms its effectiveness, while ablation studies validate the individual contribution of each component, establishing a promising and reliable direction for jailbreak detection.

\section{Limitation and Discussion}

We observed several interesting phenomena in the construction of the $\mathbb{RSD}$. Specifically, $\mathbb{RSD}$ sampling involves testing models with simple hazardous prompts under varying temperature settings and collecting their refusal patterns across different queries. We found that the collected samples vary substantially in length, as models sometimes provide partial responses to the safe portions of a query even after issuing a refusal. Consequently, certain non-refusal segments must be manually removed during data collection. However, this manual sampling process inevitably limits the generalizability of the approach to some extent.

In addition, our current experiments are conducted with only two LLM models and corresponding datasets. However, further validation across a broader range of models and datasets is necessary to rigorously verify the effectiveness and accuracy of the proposed approach. Real-world prompts should also be examined and validated, although this becomes increasingly challenging as model alignment on safety continues to improve.Many prompts and datasets that previously confused LLMs are gradually losing their effectiveness as the models become more capable. The ablation study can be extended to a more fine-grained analysis of how individual samples within the RSD affect overall performance, as well as to assess whether the salient sentence extraction algorithm consistently identifies the most representative and accurate sentence components across different response modes. Finally, although this framework achieves substantial improvements in accuracy and robustness, its efficiency remains limited in terms of resource utilization and runtime. Because it applies $k$-means clustering and Isolation Forest–based outlier detection to the responses generated for each prompt, it incurs nontrivial time and GPU overhead. Addressing these computational costs is an important direction for future work.

\bibliographystyle{IEEEtran}
\bibliography{reference}

\end{document}